\documentclass[aps,prl,twocolumn,groupedaddress]{revtex4-1}
\usepackage{amsmath}           
\usepackage{graphicx}          
\usepackage{kotex}             
\usepackage{dcolumn}           
\usepackage{here}
\usepackage[utf8]{inputenc}
\usepackage{listings}
\usepackage{lipsum}
\usepackage{subfig}

\usepackage{algorithm}
\usepackage{algpseudocode}
\usepackage{listings} 
\makeatletter
\newenvironment{breakablealgorithm}
{
	\begin{center}
		\refstepcounter{algorithm}
		\hrule height.8pt depth0pt \kern2pt
		\renewcommand{\caption}[2][\relax]{
			{\raggedright\textbf{\ALG@name~\thealgorithm} ##2\par}%
			\ifx\relax##1\relax 
			\addcontentsline{loa}{algorithm}{\protect\numberline{\thealgorithm}##2}%
			\else 
			\addcontentsline{loa}{algorithm}{\protect\numberline{\thealgorithm}##1}%
			\fi
			\kern2pt\hrule\kern2pt
		}
	}{
		\kern2pt\hrule\relax
	\end{center}
}
\makeatother

\begin{document}               

\title{Multi-Gaussian fitting Algorithm to determine multi-band photometry
	and photometric redshifts of LABOCA and Herschel sources in proto-cluster
	environments}       
 
\author{Youngik Lee}		    
\affiliation{Dept. of Physics, University of Bonn, Bonn-Cologne Graduate School, Germany}



\begin{abstract}
This research focuses on identifying high redshift galaxies from LABOCA(LArge APEX BOlometer CAmera) and SPIRE(The Spectral and Photometric Imaging Receiver) maps towards proto-cluster candidates initially selected from the SPT (South pole telescope) survey.
Based on the Multi-Gaussian fitting algorithm, we cross-match all significant LABOCA sources at SPIRE wavelengths based on their coordinates and signal to noise ratio to derive their photometry at 250, 350, 500 and 870 $\mu m$. We use this information to calculate a photometric redshift for SPT sources towards cluster fields. The code was developed in the Python programming environment.
\end{abstract}

\maketitle			        
\tableofcontents

\section{1. Introduction}

There are two main sub-millimeter observation data used in this project, which are called LABOCA and SPIRE image.\cite{1}

First, the LABOCA is a short form of "Large APEX BOlometer CAmera," which composed of 295 bolometers operating at a wavelength of 870 $\mu$m. The LABOCA is a facility instrument of APEX, and it is highly sensitive detectors that enable us to measure faint submillimeter signals received from cold, dusty astronomical sources.  
The Atacama Pathfinder Experiment Telescope (APEX) is located at 5100 meters altitude, high on the Chajnantor Plateau in Chile's Atacama region. This 12-meter telescope operating at millimeter and submillimetre wavelengths allows astronomers to study the cold universe ( gas, dust, and celestial objects) that are only a few degrees above absolute zero kelvin and the LABOCA spatial resolution is about 18.6 arcsec.

Second, the SPIRE is a short form of "Spectral and Photometric Imaging Receiver," which is one of the instruments onboard on Herschel Space Observatory (2010).
Herschel Space Observatory was a space observatory built by the European Space Agency (ESA) and was launched in 2009 from Kourou, French Guiana, and performed observations until 2013. Herschel was the largest infrared telescope ever launched, carrying a 3.5-meter mirror and instruments sensitive to the far-infrared and submillimetre wavebands.
The three major instruments were onboard in the satellite, the Heterodyne Instrument for Far Infrared (HIFI, 2010), the Photodetector Array Camera and Spectrometer (PACS, 2010) and the Spectral and Photometric Imaging Receiver (SPIRE, 2010) which performed photometry and spectroscopy observations in the infrared and the far-infrared domains, from 55 $\mu$m to 672 $\mu$m.
This spectral domain covers the cold and the dusty universe, which means from dust-enshrouded galaxies at cosmological distances down to scales of stellar formation, planetary system bodies, and our solar system objects.
In this work, we used photometric SPIRE maps at 250, 350, 500 $\mu m$n wavelength. Moreover, the SPIRE's spatial resolution is 35, 24, and 17 arcseconds at 500, 350, and 250 microns, respectively.

The purpose of this research project is to extract the relevant astronomical information, including their coordinates(RA, DEC), Intensity(mJy), uncertainties for each variable, and red-shift from the original LABOCA/SPIRE maps by using "Multiple Gaussian fitting Method." The code development environment is Python, and for error reduction, the data smoothing process was used in the LABOCA/SPIRE map, respectively, by using MIRIAD. \cite{2}

\section{2. Programming Task and Results}

\subsection{2.1 Finding the source with Multi-Gaussian peak}

The figures in \textbf{Section 2.1} show the process of how the python code finds the relevant source form the LABOCA image. The used data exemplify the different steps based on the Laboca map of \texttt{SPT0303-59}. 

For the LABOCA and SPRIE maps, we compute signal-to-noise (SNR) maps, which all analysis are done on, except the flux determination of each source. The emission detected by LABOCA and SPIRE is much smaller than the beam. Therefore we fix the FWHM in the Gaussian fitting to the beam size.

Because the emission detected with LABOCA is coming from a small region in the maps, the first step in the analysis is to automatically select smaller sub-regions (called boxes in the following), which contain the significant emission. This step is required because it limits the area used for the fitting of the sources and thereby improves the reliability and performance of the Multi-gaussian fitting.
In Fig. 1 we show the result of boxing out the source from the original global flux map, and then localized the dense source from the voids, as one can see in Fig. 2.

The Multi-Gaussian fitting algorithm is first to find the highest SNR peak, and next fit a single Gaussian to it and subtract the fit from the SNR data. This process is repeated until all SNR values fall below the significance threshold, which we set to 3.0 sigma. In each step, we record the x,y positions and the SNR peak value from the single Gaussian fit as an initial guess for the Multi-Gaussian fitting.

When after all subtraction process is finished, then for the final stage, sum up all the peak information (coordinate, flux density, SNR, uncertainties for each variable) and fitted simultaneously with the original data by using Multi-Gaussian fitting (Fig. 4(d)).
The region we can see in Fig.3 is one of the boxed range from the sky map. It shows a zoom of the largest, continues emission region which the origin point is from coordinate (138, 146) in Fig. 2 (The box with three source peaks).

\begin{figure}[H]
	\centering
	\includegraphics[angle=0, width=8.5cm, height=6cm]{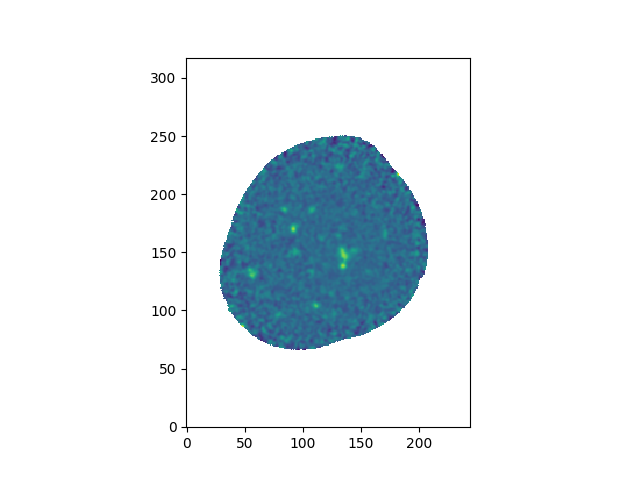}
	\caption{\label{fg3} original global flux map}
\end{figure}

\begin{figure}[H]
	\centering
	\includegraphics[angle=0, width=8.5cm, height=6cm]{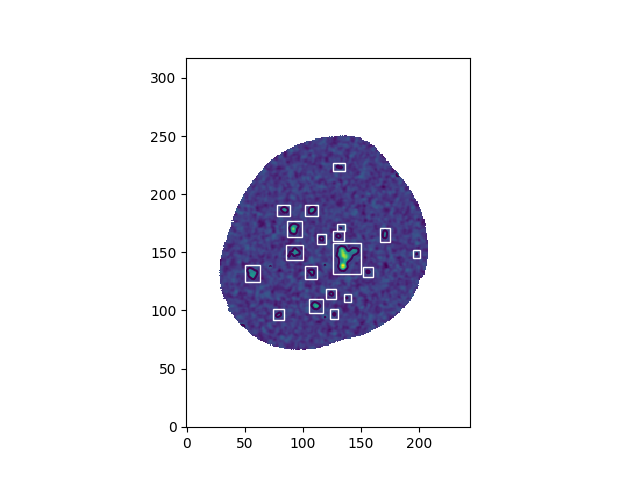}
	\caption{\label{fg3} The boxed area result of global flux map}
\end{figure}

\begin{figure}[H]
	\centering
	\includegraphics[angle=0, width=8.5cm, height=6cm]{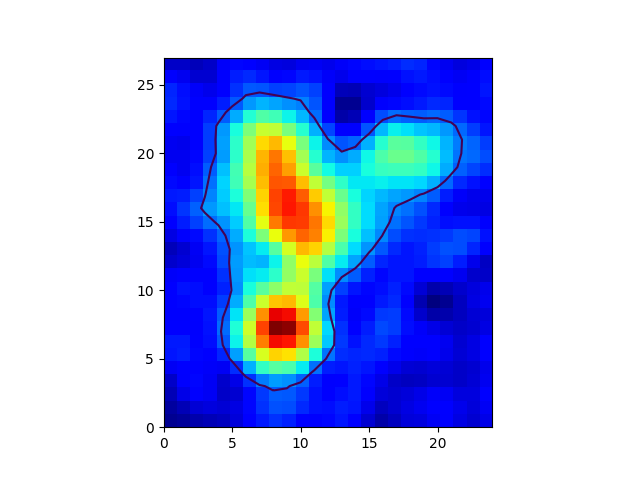}
	\caption{\label{fg3} The one of the boxed area result of global flux map, the spatial axis are pixel coordinated from the (138, 146) point of Fig. 2}
\end{figure}

\begin{breakablealgorithm}
	\caption{One-point Gaussian function code}
	\label{EPSA}
\begin{lstlisting}[linewidth=\columnwidth,breaklines=true,language=Python]
	def g1((x, y), amplitude, xo, yo, sigma):
		xo = float(xo)
		yo = float(yo)
		g = amplitude*np.exp( -1./2.*((x-xo)**2+(y-yo)**2)/sigma**2)
	return g.ravel()
\end{lstlisting}
\end{breakablealgorithm}

As we can see in the \textbf{Algorithm.1}, basically one-point Gaussian function needs coordinates (x,y) and the initial values. Moreover, for the fitting algorithm, we put the initial value as the highest peak information of the original data map.

\begin{breakablealgorithm}
	\caption{One-point Gaussian fitting}
\begin{lstlisting}[linewidth=\columnwidth,breaklines=true,language=Python]
	def fit_g1(data='',guess=[0.005,12.0,40.0], fwhm=20.,postol=0.01,stol=0.1)
		ny,nx=data.shape
		for(i to range(ny))
		for(j to range(nx))
		if{np.isnan(data[i][j])}:
			data[i][j] = 0.0
			x = np.linspace(0, nx-1, nx)
			y = np.linspace(0, ny-1, ny)
		x, y = np.meshgrid(x, y)
		myguess=guess[:3]
		sigma=fwhm/(2.*np.sqrt(2.*np.log(2.)))
		myguess.append(sigma)
		mybounds=([0,guess[1]-postol,guess[2]-postol,sigma-stol*sigma],[np.inf,guess[1]+postol,guess[2]+postol,sigma+stol*sigma])
		popt, pcov = $opt.curve_fit$(g1, (x, y), data.ravel(), p0=myguess,bounds=mybounds)
		data_fitted = g1((x, y), *popt)
	return data_fitted.ravel(), popt, pcov
\end{lstlisting}
\end{breakablealgorithm}

In \textbf{Algorithm. 2}, \texttt{opt.curve\_fit} is the main fitting function that can use in the Python library. Moreover, for using this function, we need to input fitting function, fitting point, and data with initial values which call \texttt{myguess}, \texttt{mybounds}.

\begin{figure*}[htp]
	\centering
	\subfloat[ The first peak subtraction from the source.  ]{\label{figur:1}\includegraphics[width=85mm]{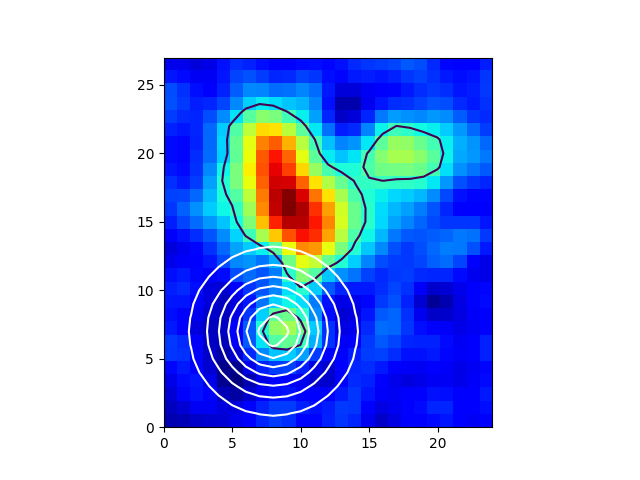}}
	\subfloat[ The second peak subtraction from source. ]{\label{figur:2}\includegraphics[width=85mm]{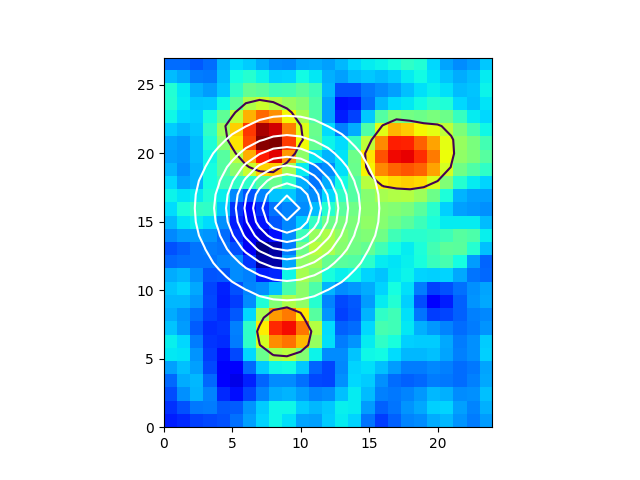}}
	\\
	\subfloat[The third peak subtraction from source.]{\label{figur:3}\includegraphics[width=85mm]{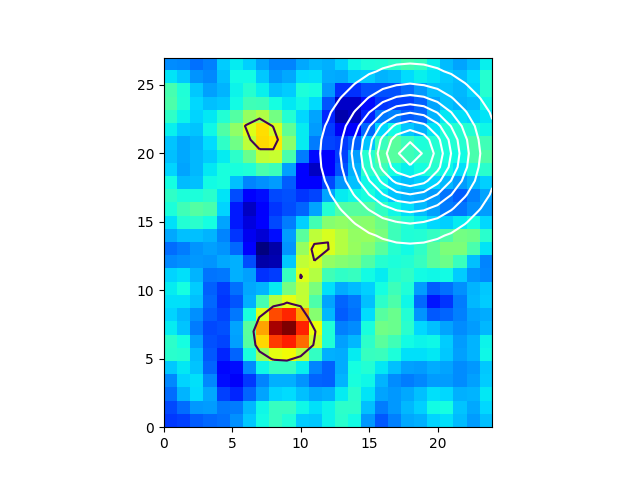}}
	\subfloat[The multiple gaussian fit after three peak subtraction from the source.]{\label{figur:4}\includegraphics[width=85mm]{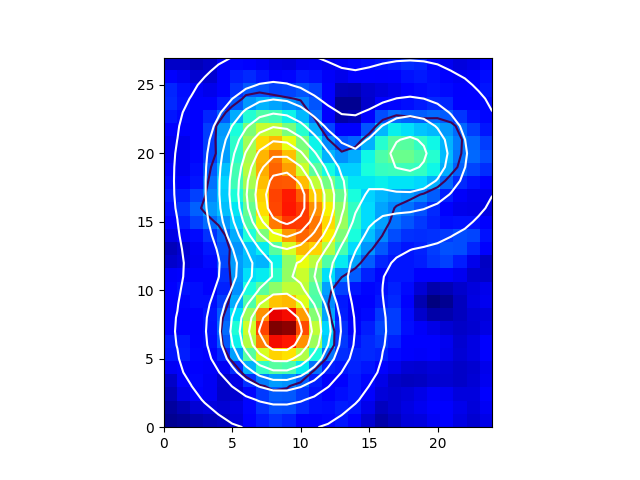}}
	
	\label{figur}\caption{The one-point Gaussian peak subtraction process. The spatial axis is pixel coordinated for the (138, 146) point of the box in Fig. 2}    
\end{figure*}

Fig. 4 (a) shows the position of highest peak values from the original data and subtracts the one-point gaussian distribution around it. (The code can be seen at \textbf{Algorithm. 2}) The algorithm is oriented to find the location of the highest peak inside the data and adjust the one-point gaussian fit at that point. Moreover, after that, subtract the gaussian estimated values from the original map.
Fig. 4 (b) shows the subtraction of the next-order-highest peak from Fig. 4 (a), under the same process. Moreover, Fig. 4 (c) also shows the result after subtracting the three highest value peaks form the original data.
After Fig. 4 (c), no pixels with an SNR value, which is larger than 3.0 are found in the residual map, and the algorithm returns the fitted SNR map and its x, y coordinates for each Gaussian as a list.

\begin{breakablealgorithm}
	\caption{Multi-Gaussian function Code}
	\begin{lstlisting}[linewidth=\columnwidth,breaklines=true,language=Python]
def F(n):
	if n==1:
		return g1
	else:
		arg=[0.0]*(4*n)
		def gn((x, y), *arg):
			g = arg[0]*np.exp( -1./2.*((x-arg[n])**2+(y-arg[2*n])**2)/arg[3*n]**2)
			for i in range(1,n):
				g += arg[i]*np.exp( -1./2.*((x-arg[n+i])**2+(y-arg[2*n+i])**2)/arg[3*n+i]**2)
		return g.ravel()
	return gn
\end{lstlisting}
\end{breakablealgorithm}

In the next step, we use this information in the multi-Gaussian fitting as an initial guess. However, before doing Multi-Gaussian fitting, we need to define the Multi-Gaussian function. (\textbf{Algorithm. 3}) which use the "function-in-function" technique. So we define a new function "\texttt{gn()}" in \texttt{F(n)}.

Also in \textbf{Algorithm. 4}, we define the \texttt{fit\_gn} function inside the \texttt{multiGaussDefine(k)} because we don't know how many peaks we can find in the box range. The basic algorithm is the same as one-point Gaussian fitting but needs an initial guess for each Gaussian to run the \texttt{opt.curve\_fit}, provided by  \textbf{Algorithm. 3}.
And by using \texttt{append} function make the \texttt{myguess} and \texttt{mybounds}.
To find out the number of Gaussian peaks in the localized source, we do the sum-up of the number of single Gaussians.

The fitting the sum of single Gaussian to the source structure is only the first step to derive the number of Gaussian components and a guess for their location and intensity. The FWHM of the Gaussian is fixed with a small margin to the beamwidth of the map that this decomposition of the source structure is then used as an initial guess for the simultaneous fitting of all gaussian components.

\begin{breakablealgorithm}
	\caption{Multi-Gaussian fitting}
\begin{lstlisting}[linewidth=\columnwidth,breaklines=true,language=Python]
def multiGaussDefine(k):
	guess=[0.0]*(3*k)
	def fit_gn(data='', fwhm=20., postol=0.01,stol=0.05,*guess):
		ny,nx=data.shape
		for i in range(ny):
			for j in range(nx):
			if np.isnan(data[i][j]):
				data[i][j] = 0.0

		x = np.linspace(0, nx-1, nx)
		y = np.linspace(0, ny-1, ny)
		x, y = np.meshgrid(x, y)
		sigma=fwhm/(2.*np.sqrt(2.*np.log(2.)))
		myguess=list(guess[:3*k])
		for n in range(k):
			myguess.append(sigma)

		mini=list(guess[:k])
		for n in range(k):
			mini[n]=0.0
		for n in range(k):
			mini.append(guess[k+n]-postol)
		for n in range(k):
			mini.append(guess[2*k+n]-postol)
		for n in range(k):
			mini.append(sigma-stol*sigma)

		maxi=list(guess[:k])
		for n in range(k):
			maxi[n]=np.inf
		for n in range(k):
			maxi.append(guess[k+n]+postol)            
		for n in range(k):
			maxi.append(guess[2*k+n]+postol)
		for n in range(k):
			maxi.append(sigma+stol*sigma)

		mybounds=(mini,maxi)
		gn=F(k)
		popt, pcov = opt.curve_fit(gn, (x, y), data.ravel(), p0=myguess, bounds=mybounds)

		data_fitted = gn((x,y),*popt)
		return data_fitted.ravel(), popt, pcov
	return fit_gn
\end{lstlisting}
\end{breakablealgorithm}

The result of the Multi-Gaussian fitting can be shown in Fig. 4 (d).
For fitting the fluxes, we use the fitting result from the signal-to-noise map to fix the coordinates. It is advantages to fit the source coordinates on the SNR map to avoid that insignificant noise peaks are considered.
For the flux fitting, the only free parameter for the multi-gaussian is the flux density. In each step (also in the SNR fitting), the residual map is analyzed to verify that no significant emission is missing for the final multi-gauss fitting, and the residual of the multi-Gaussian fitting is finally rechecked.

\subsection{2.2 The flux, fitted, residual fits for Global map}

In the next step, we run the Multi-Gaussian fitting technique on each box of the map to obtain the coordinates, the SNR and the flux density from all significant sources.
Fig. 5 (a) shows you the original LABOCA image (\texttt{Laboca/SPT0303-59}).

\begin{figure*}[htp]
	\centering
	\subfloat[ The original flux map.]{\label{figur:1}\includegraphics[width=85mm]{5.png}}
	\subfloat[ The boxed map ]{\label{figur:2}\includegraphics[width=85mm]{21.png}}
	\\
	\subfloat[The source extracted map by using gaussian fit. ]{\label{figur:3}\includegraphics[width=85mm]{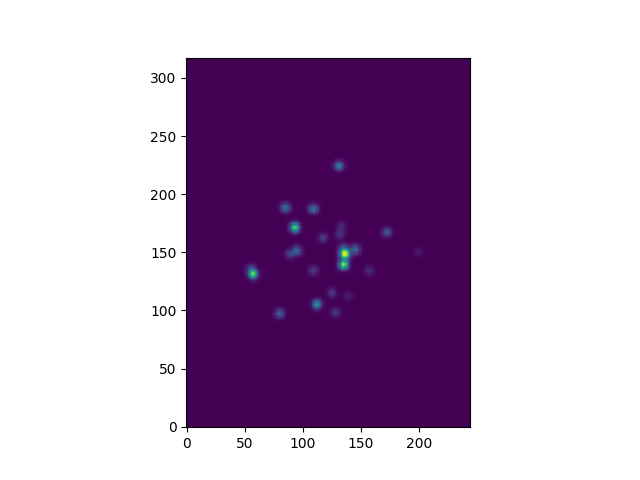}}
	\subfloat[The residual map ]{\label{figur:4}\includegraphics[width=85mm]{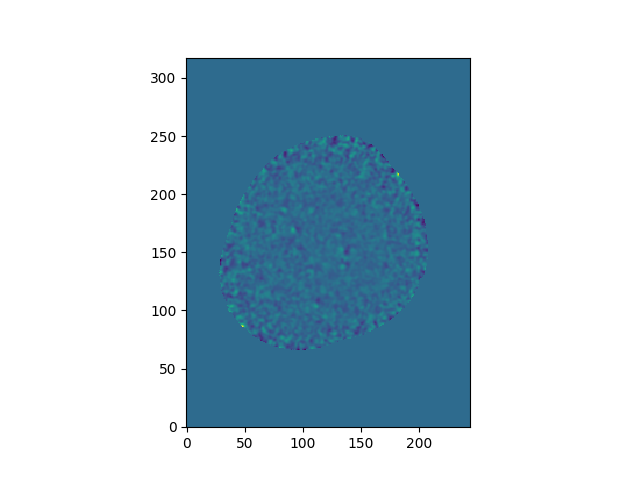}}
	\label{figur}\caption{Example for the source extraction on a LABOCA map for SPT0303-59.  Compare to the original flux map (Fig. 5 (a)), in Fig. 5 (c) shows the bright peaks are decomposed into individual galaxies from Figure (b)}    
\end{figure*}

In Fig. 5 (b) we show the result of the multi-Gaussian fitting. As we mentioned in \textbf{Section 2.1}, we adjust multi-gaussian fitting in all boxed and replot it on the whole map. And in Fig. 5 (d), we show the residual map obtained by subtracting the fitting result (Fig. 5 (c)) from the original map (Fig. 5 (a)). No significant emission with SNR $>$ 3.0 is present in the map after the multi-Gaussian fitting.

For the Herschel image,  in Fig. 6 we show the source extraction for the (\texttt{SPT2052-62}, for the SPIRE 350$\mu m$ map.  
In general, the sky coverage of the SPIRE images differs from the LABOCA maps.
In the SPIRE maps, we detect many more galaxies than in the LABOCA map ordinarily because galaxies are brighter as shorter wavelengths. This implies that SPIRE also detects a lot of foreground galaxies which remain undetected by LABOCA. The cross-matching between different wavelength is described in the \textbf{Section 2.3}

\begin{figure*}[htp]
	\centering
	\subfloat[The flux map of the Herschel image.]{\label{figur:1}\includegraphics[width=85mm]{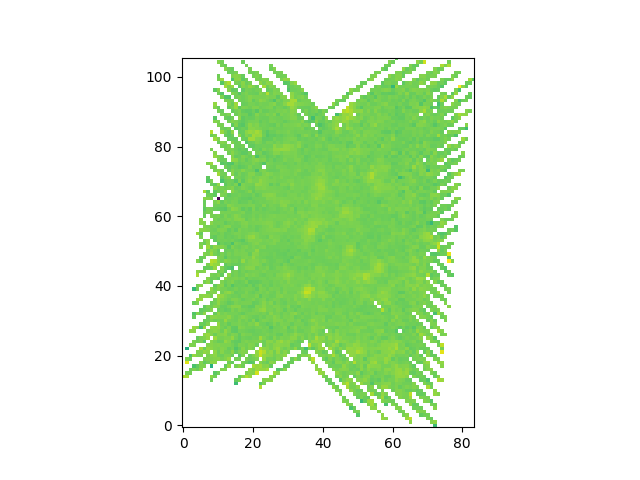}}    
	\subfloat[The boxed map of Herschel image]{\label{figur:2}\includegraphics[width=85mm]{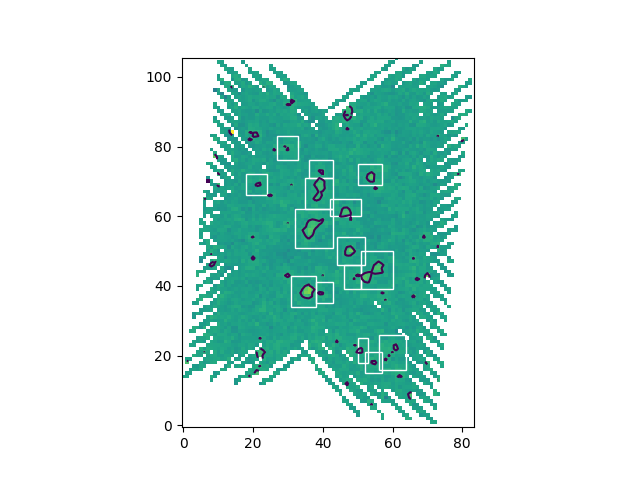}}
	\\
	\subfloat[The source extracted map by using gaussian fit]{\label{figur:3}\includegraphics[width=85mm]{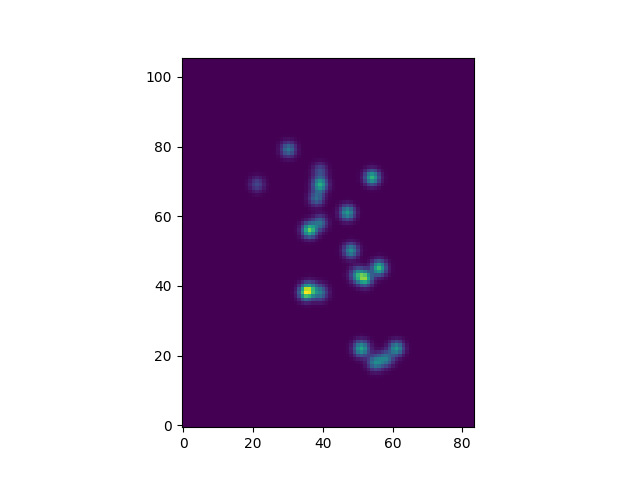}}    
	\subfloat[The residual map. ]{\label{figur:4}\includegraphics[width=85mm]{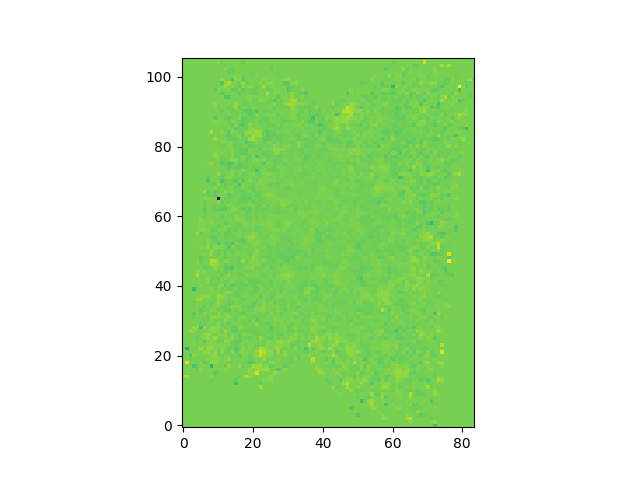}}
	
	\label{figur}\caption{Example for the source extraction on a SPIRE map for SPT2052-62 at 350 $\mu m$. The individual panels are the same as in Fig. 5 }    
\end{figure*}

Fig. 6 (a) shows you the original SPIRE image we need to which we aim to decompose. Moreover, in Fig. 6 (c), you can see the fitted result for multi-Gaussian fitting. Compare to the LABOCA image, it covers more small range, but we can see more detail distributions of a galaxy with other wavelength scales.
And in Fig. 6 (d), we can see the residual map, which is subtracted the fitted result from the original map, subtract Fig. 6 (c) from Fig. 6 (a)

\subsection{2.3 Extracting the source information}

After extracting all source from each map, we save the equatorial world map coordinates in the J2000, SNR, flux density, and the uncertainty of each parameter, in \texttt{.txt} file.

\begin{breakablealgorithm}
	\caption{Making sourcelist code}
	\begin{lstlisting}[linewidth=\columnwidth,breaklines=true,language=Python]
## PRINT RESULT AND ADD SOURCES TO SOURCE LIST 
print "Fit result"
for i in range(k):
wx, wy = w.wcs_pix2world(posx[i],posy[i],0)
print "%i peak SNR: %6.2f +/- %6.2f peak Flux: %6.2f +/- %6.2f [mJy]  X: %20.16f  Y: %20.16f" %(i+1, snr[i],dsnr[i],1000.*flux[i],1000.*dflux[i],wx,wy)

if snr[i] > snrcut and dsnr[i]<snr[i] and dflux[i]<flux[i]:
## ADD SOURCE TO SOURCE LIST
sourcelist[0].append(np.float(wx))
sourcelist[1].append(np.float(wy))
sourcelist[2].append(snr[i])
sourcelist[3].append(dsnr[i])
sourcelist[4].append(flux[i])
sourcelist[5].append(dflux[i])
## ADD SOURCE TO THE FLUX MODEL
fluxmodel = fluxmodel+g1((xorg,yorg),flux[i],posx[i],posy[i],mysig).reshape(nyorg,nxorg)
	\end{lstlisting}
\end{breakablealgorithm}

One can see in \textbf{Algorithm. 5} lines 9 to 14 describes saving the coordinate, SNR, and flux density with relevant errors. And in  \textbf{Algorithm. 6} is converting the source list data into \texttt{.txt} files.

\begin{breakablealgorithm}
	\caption{Save sourcelist as \texttt{txt} file}
	\begin{lstlisting}[linewidth=\columnwidth,breaklines=true,language=Python]
	## WRITE ASCII FILE WITH SOURCE CATALOG
if "SPIRE" in infile:
outtxt=infile.split(".fits")[0]+".txt"
else:
outtxt=infile.split(".fits")[0]+"_LABOCA.txt"

f=open(outtxt,'w')
f.write("#Index    RA            DEC          SNR      dSNR   FLUX [mJy]  dFLUX [mJy] \n")
for i in range(len(sourcelist[0])):
if sourcelist[2][i] > snrcut and (sourcelist[3][i]<=sourcelist[2][i] and sourcelist[5][i]<=sourcelist[4][i]):
s=Deg2HMS(np.float(sourcelist[0][i]),np.float(sourcelist[1][i]))
ra=s.split(" ")[0].split(".")[0]+"."+s.split(" ")[0].split(".")[1][:2]
dec=s.split(" ")[1].split(".")[0]+"."+s.split(" ")[1].split(".")[1][:1]
f.write("%3i	%s	%s	%6.2f	%6.2f	%6.2f	%6.2f\n"% (i+1,ra,dec,sourcelist[2][i],sourcelist[3][i],1000.*sourcelist[4][i],1000.*sourcelist[5][i]))
f.close()
	\end{lstlisting}
\end{breakablealgorithm}

After saving the source list for all LABOCA/SPIRE map, we can sort out the source which contains relevant flux density of all four wavelengths (250, 350, 500, 850 $\mu m$).
Cross-matching the galaxies identified in the LABOCA map on the SPIRE bands is a multi-step process. First, we compare the coordinates in an individual tolerance, secondly do gaussian-fitting at the Laboca position on the SPIRE map. And lastly integrating the pixel fluxes at the LABOCA position in case the
Gaussian fit does not give reliable results. 
Because the source may not show up in the SPIRE maps (in particular at 350 or 250$\mu m$) and to verify that whether the flux densities are extracted belong to the same object.

\subsection{2.4 Save the fitted/Residual data as \texttt{.fits} file}

We save the flux model map as well as the residual map for each source and wavelength for further analysis as a \texttt{fits} file.
Save the fitted map and residual map in \texttt{.fits} format by using the "\texttt{astropy.io}" package.
\textbf{Algorithm. 7} shows the algorithm for \texttt{.fit} file saving process and in \textbf{Algorithm. 8} we can see explicit code for the \texttt{writeFits} function. 

\begin{breakablealgorithm}
	\caption{\texttt{wrtieFits} function code}
	\begin{lstlisting}[linewidth=\columnwidth,breaklines=true,language=Python]
## WRITE FLUX MODEL AND RESIDUAL TO FITS FILE
outfile = infile.split(".fits")[0]+"_fitted.fits"
writeFits(infile,outfile,fluxmodel)
outfile = infile.split(".fits")[0]+"_residual.fits"
residual = fluxData-fluxmodel
writeFits(infile,outfile,residual)
	\end{lstlisting}
\end{breakablealgorithm}

In \textbf{Algorithm. 7}, the \texttt{writeFits} erases the unnecessary layers from the original \texttt{.fit} file and substitutes it with new data (e.g., residual, fitted maps)

\begin{breakablealgorithm}
	\caption{\texttt{wrtieFits} function code}
	\begin{lstlisting}[linewidth=\columnwidth,breaklines=true,language=Python]
def writeFits(infile, outfile, mydata):
newFile=fits.open(infile)
hdr = fits.getheader(infile)
primary_hdu = fits.PrimaryHDU(header=hdr)

hdu =newFile[1]
hdu.data=mydata
hdul = fits.HDUList([primary_hdu,hdu])
com = "rm -rf "+outfile
os.system(com)
hdul.writeto(outfile) 
	\end{lstlisting}
\end{breakablealgorithm}

\subsection{2.5 Checking \texttt{.fits} data with \texttt{kvis}}

To verify the source extraction and the flux models, we use the \texttt{kvis} program for visualization.
This program is part of the \texttt{KARMA} visualization packaged to form the ATNF(Australia Telescope National Facility).
In Fig. 7 we show as an example the flux model of \texttt{Laboca/SPT0303-59} at 870 $\mu m$ (contours) overlayed with the observed flux map. Moreover, we can see that the peak is correct with the source position.
Additional information (such the source ID and its position) can be displayed in \texttt{kvis} using an annotation file. We generate these \texttt{.ann} files from the source information stored in the \texttt{.txt} file.
Using different colors for different wavelengths allows us to easily display the source extraction results in each map to verify our cross-matching result.

\textbf{Algorithm. 9} shows a general process for data saving. The critical step for this is decomposing into individual galaxies source data from the source list, which error is less than its intensity, at line 20

\begin{figure}[H]
	\centering
	\includegraphics[angle=0, width=8cm, height=5cm]{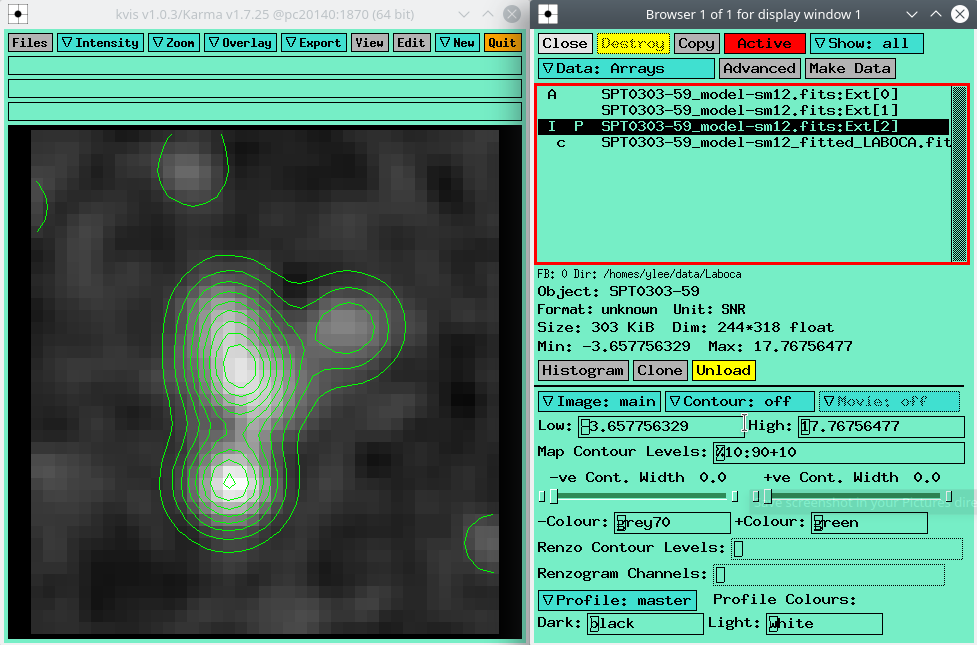}
	\caption{\label{fg3} The \texttt{kvis} image, we can check the specific source position by comparing with bare eyes.}
\end{figure}

\begin{breakablealgorithm}
	\caption{Save sourcelist as \texttt{ann} file}
	\begin{lstlisting}[linewidth=\columnwidth,breaklines=true,language=Python]
## WRITE KARAM annotation file
symbsize = 0.0040
if "SPIRE" in infile:
	outann=infile.split(".fits")[0]+".ann"
else:
	outann=infile.split(".fits")[0]+"_LABOCA.ann"
	
f=open(outann,'w')
if "LABOCA" in outann:
	f.write("color red\n")
if "SPIRE500" in outann:
	f.write("color yellow\n")
if "SPIRE350" in outann:
	f.write("color green\n")
if "SPIRE250" in outann:
	f.write("color blue\n")
	
f.write('FONT hershey22\n')
for i in range(len(sourcelist[0])):
	if sourcelist[2][i] > snrcut and (sourcelist[3][i]<=sourcelist[2][i] and sourcelist[5][i]<=sourcelist[4][i]):
	f.write("cross %20.16f %20.16f %6.5f %6.5f #%i\n"% (sourcelist[0][i],sourcelist[1][i],symbsize, symbsize*np.cos(np.float(sourcelist[1][i])*np.pi/180.), i+1))
	f.write("text W %20.16f %20.16f %i\n"% (sourcelist[0][i]+20./3600.,sourcelist[1][i], i+1))
f.close()
\end{lstlisting}
\end{breakablealgorithm}

\subsection{2.6 Plotting red-shift graph}

After verifying our cross-matching results manually, the final step in the analysis is to determine the photometric redshift for all LABOCA sources using our flux determinations at 870, 500, 350, and 250 $\mu m$.
For the redshifts, we fit a modified black spectrum to all four photometric data points for each galaxy. 
For calculating the redshift, the dust temperature is degenerate parameters, so to solve the redshift, we have to assume a dust temperature.
In practice, we use the SPT sources with known redshift from spectroscopy and derive their dust temperature. This gives a roughly Gaussian distribution with a mean of 35K and a dispersion of 8K.
This dust temperature distribution is used to solve for the redshift for the sources studied here.

\begin{breakablealgorithm}
	\caption{redshift graph printing code}
	\begin{lstlisting}[linewidth=\columnwidth,breaklines=true,language=Python]
def graph():
	for source in sourcelist:
	infile1=str(source)+"-graph.txt"
	infile2=str(source)+"-graph-sm.txt"
	f=open(infile1,'r')
	lines=f.readlines()
	f.close()
	f2=open(infile2,'r')
	lines2=f2.readlines()
	f2.close()
	lobs1=[1,0,0,0]; sobs1=[1,0,0,0]; dsobs1=[1,0,0,0]
	lobs2=[1,0,0,0]; sobs2=[1,0,0,0]; dsobs2=[1,0,0,0]
	for i in range(len(lines)/4):
	for k in range(len(lines2)/4):
	if lines[4*i].split(' ')[1].split('_')[1]==lines2[4*k].split(' ')[1].split('_')[1]:
	title = lines[4*i]; title2 = lines2[4*k]
	for j in range(4):
	print source, lines[4*i+3].split('[')[1].split(']')[0].split(',')
	lobs1[j] = float(lines[4*i+1].split('[')[1].split(']')[0].split(',')[j])
	sobs1[j] = float(lines[4*i+2].split('[')[1].split(']')[0].split(',')[j])
	dsobs1[j]= float(lines[4*i+3].split('[')[1].split(']')[0].split(',')[j])
	lobs2[j] = float(lines2[4*k+1].split('[')[1].split(']')[0].split(',')[j])
	sobs2[j] = float(lines2[4*k+2].split('[')[1].split(']')[0].split(',')[j])
	dsobs2[j]= float(lines2[4*k+3].split('[')[1].split(']')[0].split(',')[j])
	snu1, pfit1, ffit1, sfit1, dsfit1 = fits(lobs1, sobs1, dsobs1)
	snu2, pfit2, ffit2, sfit2, dsfit2 = fits(lobs2, sobs2, dsobs2)
	if np.min(ffit1) < 100:
	flow1 = np.min(ffit1)
	else:
	flow1 = 100.
	if np.max(ffit1) >1800:
	fhigh1 = np.max(ffit1)
	else:
	fhigh1 = 1800
	f1=np.array(range(np.int(flow1),np.int(fhigh1)))
	if np.min(ffit2) < 100:
	flow2 = np.min(ffit2)
	else:
	flow2 = 100.
	if np.max(ffit2) >1800:
	fhigh2 = np.max(ffit2)
	else:
	fhigh2 = 1800
	f2=np.array(range(np.int(flow2),np.int(fhigh2)))
	x1 = ffit1; x2 = ffit2; y1 = sfit1; y2 = sfit2	

	## plotting
	plt.clf()
	plt.semilogx()
	plt.semilogy()
	plt.plot(f1,snu1(pfit1,f1),'k',label='Fits, '+str(lines[4*i].split(' ')[2])+str(lines[4*i].split(' ')[3].split('\n')[0]))
	plt.plot(f2,snu2(pfit2,f2),linestyle='solid',label='Fits-smoothed, '+str(lines2[4*k].split(' ')[2])+str(lines2[4*k].split(' ')[3].split('\n')[0]))
	plt.plot(x1,y1,'ro',color='r',label='raw')
	plt.plot(x2,y2,'ro',color='b',label='smoothed')
	plt.legend(loc='best')
	plt.title(title,loc='center')
	plt.xlabel("Frequency [GHz]")
	plt.ylabel("Intensity [mJy]")
	plt.savefig(title, format='eps')
\end{lstlisting}
\end{breakablealgorithm}

When after getting all residual, extracted map for all LABOCA and Herschel maps, now we can run all the processes related to getting redshift for each source. In \textbf{Algorithm. 10}, there is a \texttt{graph()} function which plot Fig. 8 to Fig. 10.
From line 1 to 45 is for setting the data, and When we see line 2 and 3, we call the \texttt{.txt} file. And in these files, there are lobs and sobs variables. The code for redshift, which is not introduced in this section, produces infile 1 and infile 2 in \textbf{Algorithm. 10} lines 3 and 4. 
Here \texttt{lobs} and \texttt{sobs} at line 11, 12 is the data frame which saves 250, 350, 500, 870 $\mu m$ intensities.
Line 47 to 59 describes the plotting process.

In Fig. 8 to Fig. 10, we show examples of the derived photometry and the fitting result of the modified black body for three sources. These plots are produces by the \texttt{graph()} function in in \textbf{Algorithm. 10}.
For the fitting of the photometric redshifts, we compare the results we derive for the source extraction performed on the spatially un-smoothed maps (raw, red data points in Fig 8 to 10) and after slightly smoothing all input maps spatially. The later improves the signal-to-noise ratio of the sources but also gives slightly different flux densities in particular at 350 and 250 $\mu m$.
A comparison between the theoretical modified black body fit and the data allows us to check the consistency of our photometric data points. In Fig. 8, we show an example for well-determined photometry (here e.g., ID SPT2052-56 7 means the 7th source of the SPT2052-56 system) where the data points follow the theoretical line we expect. 
For this source, we derive a redshift of 2.5$\pm$0.5 from the flux extraction on the unsmoothed maps and a slightly lower redshift of 1.8$\pm$0.3  from the smoothed maps. Typical uncertainties for the photometric redshifts are ± 0.5. The uncertainty increase in case of less consistent photometry (see e.g., Fig 9 and 10).

\begin{figure}[H]
	\centering
	\includegraphics[angle=0, width=9.5cm, height=6.5cm]{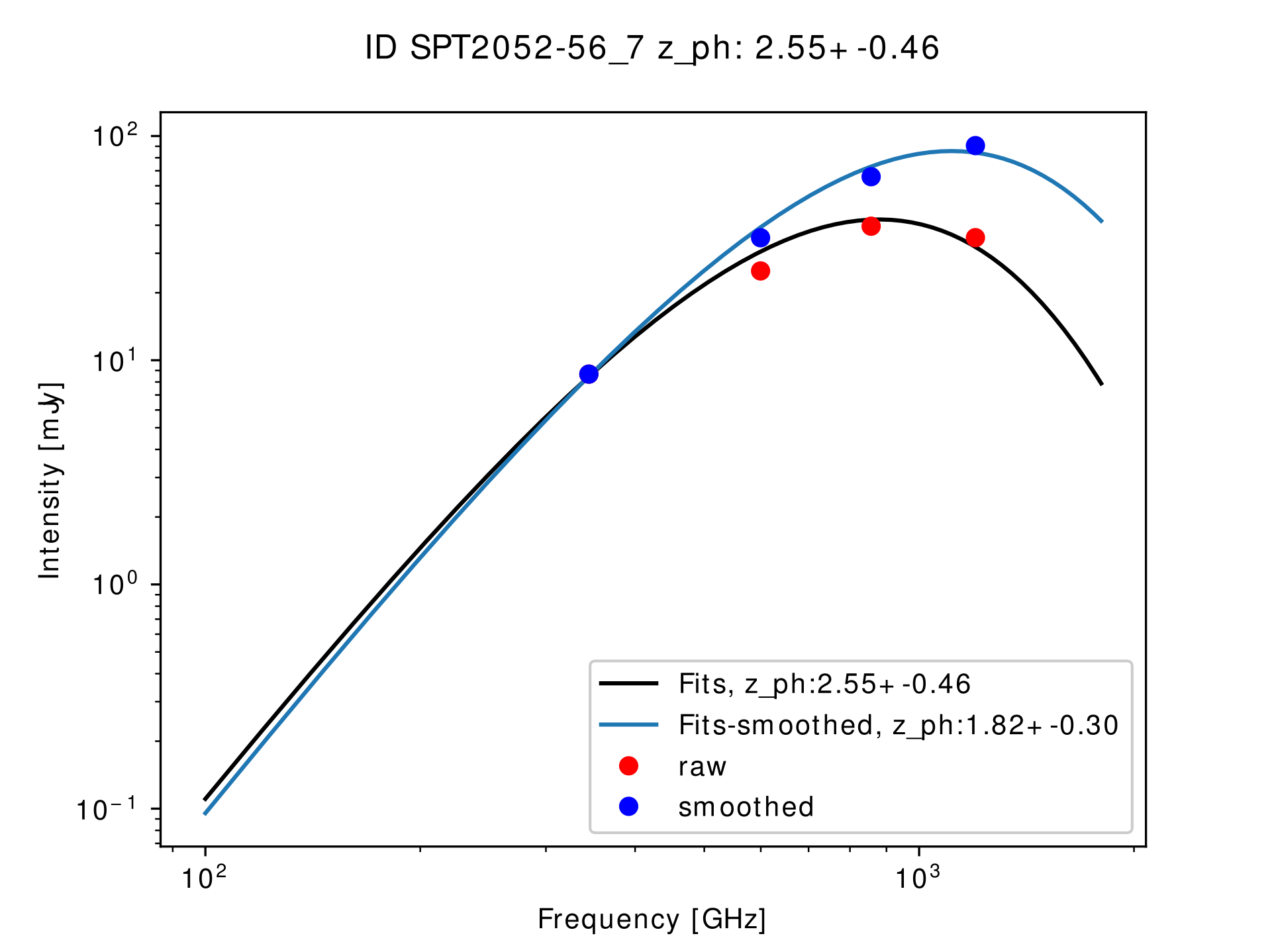}
	\caption{\label{fg3} \texttt{ID SPT2052-56\_7}, $z_{ph}: 2.55\pm0.46$, smoothed version also gives similar accuracy with original result. But in smoothed version the red-shift are increased almost twice larger}
\end{figure}

\begin{figure}[H]
	\centering
	\includegraphics[angle=0, width=9.5cm, height=6.5cm]{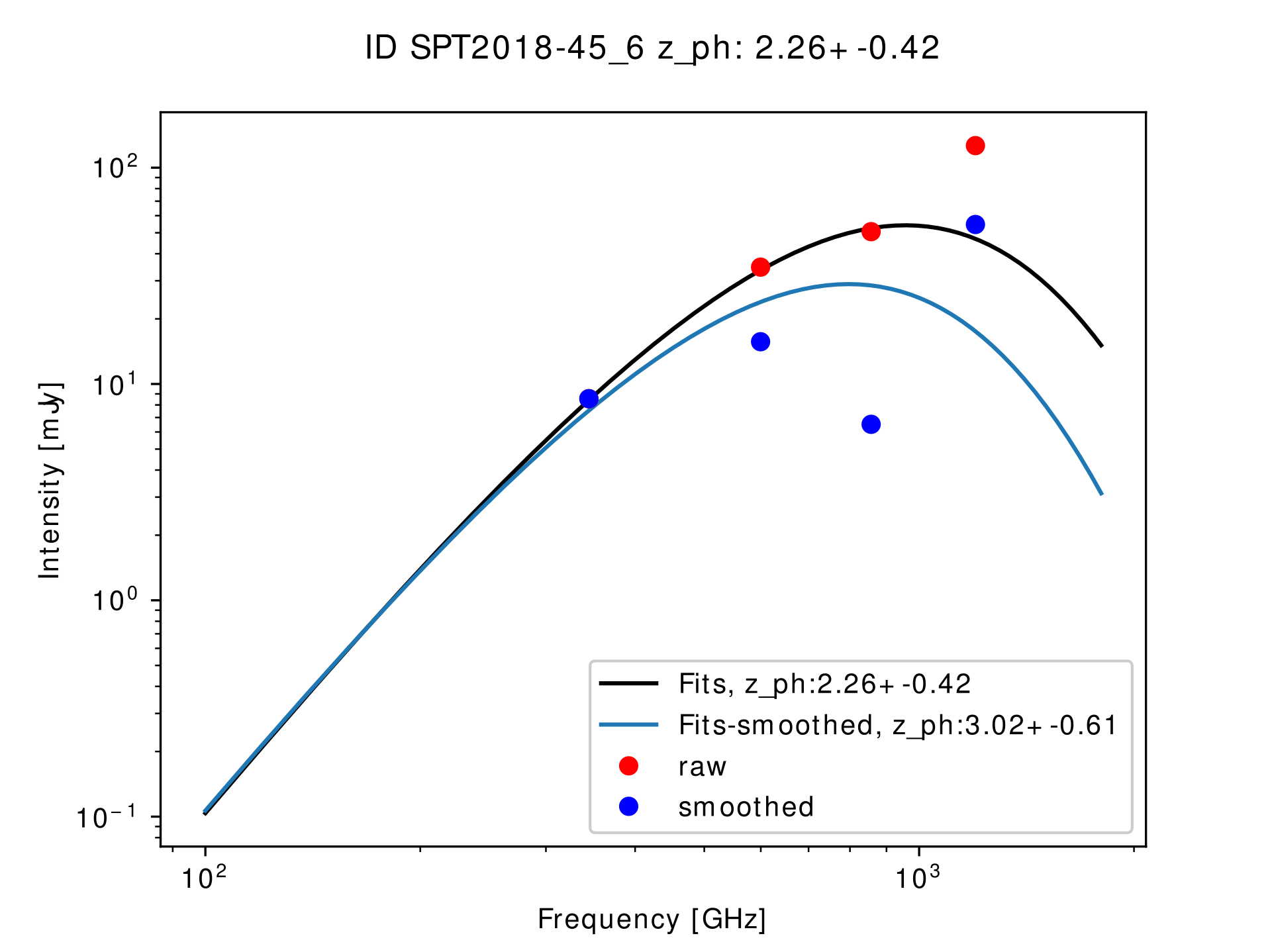}
	\caption{\label{fg3} \texttt{ID SPT2018-45\_6}, $z_{ph}: 2.26\pm0.42$, In smoothed version, around 1000 Hz frequency region it gives reasonable modified result for original curve}
\end{figure}

\begin{figure*}[htp]
	\centering
	\subfloat[\texttt{ID SPT0303-59\_12}, $z_{ph}: 2.32\pm0.36$]{\label{fg3}\includegraphics[width=8cm]{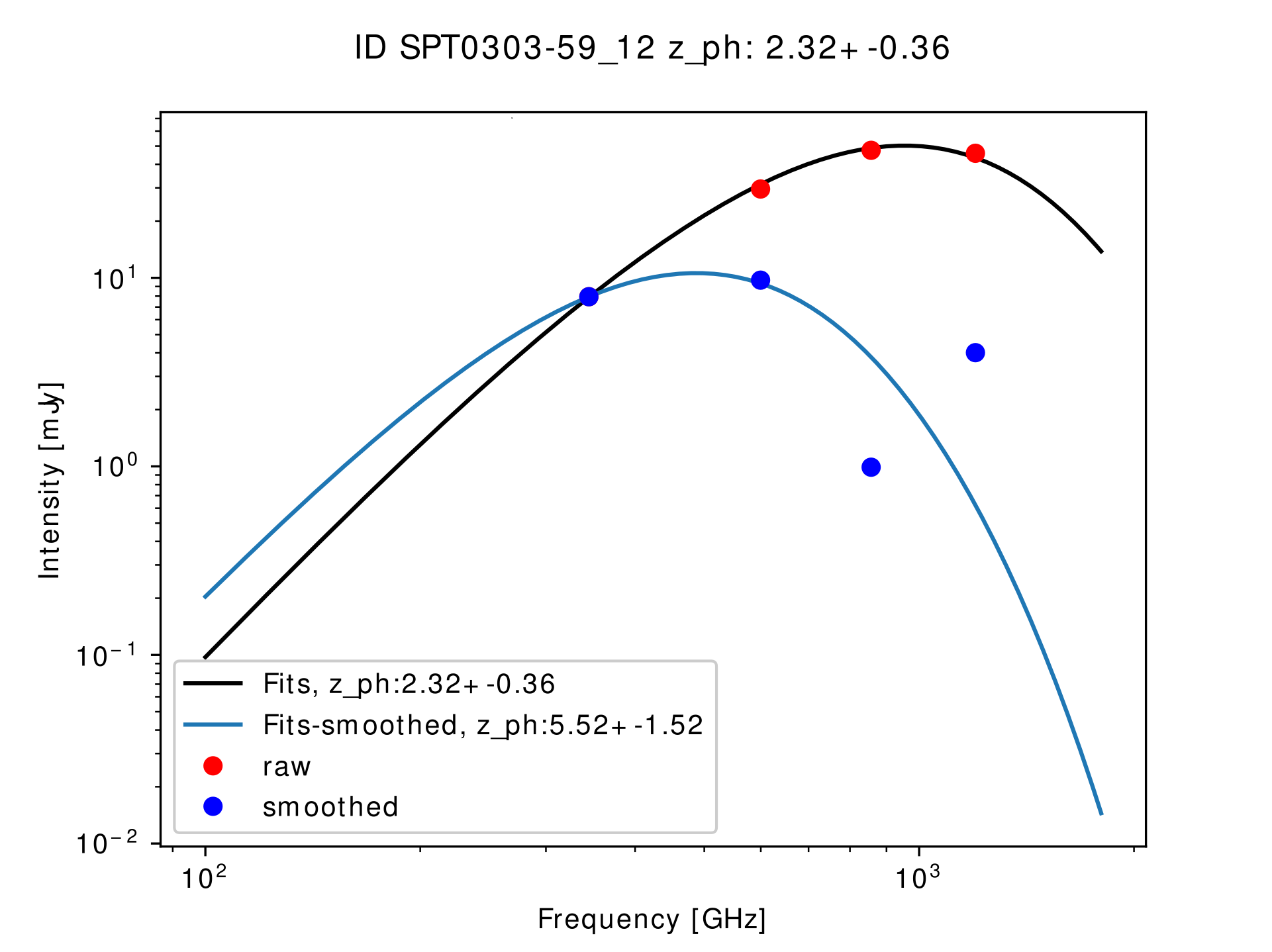}}
	\subfloat[\texttt{ID SPT0533-50\_1}, $z_{ph}: 5.23\pm1.29$]{\label{fg3} \includegraphics[width=8cm]{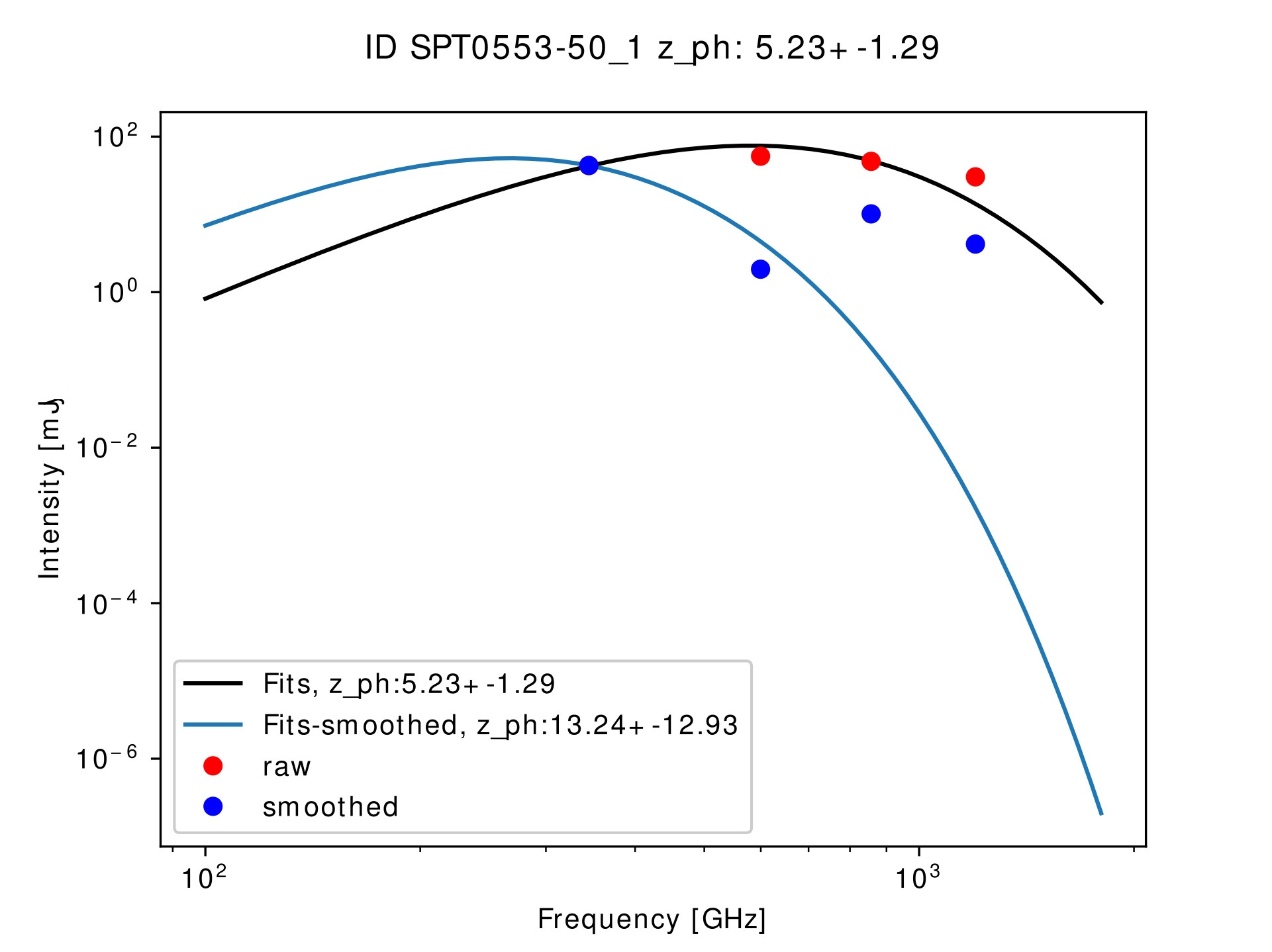}}
	\\	
	\subfloat[\texttt{ID SPT2052-56\_6}, $z_{ph}: 4.04\pm0.82$]{\label{fg3}\includegraphics[width=8cm]{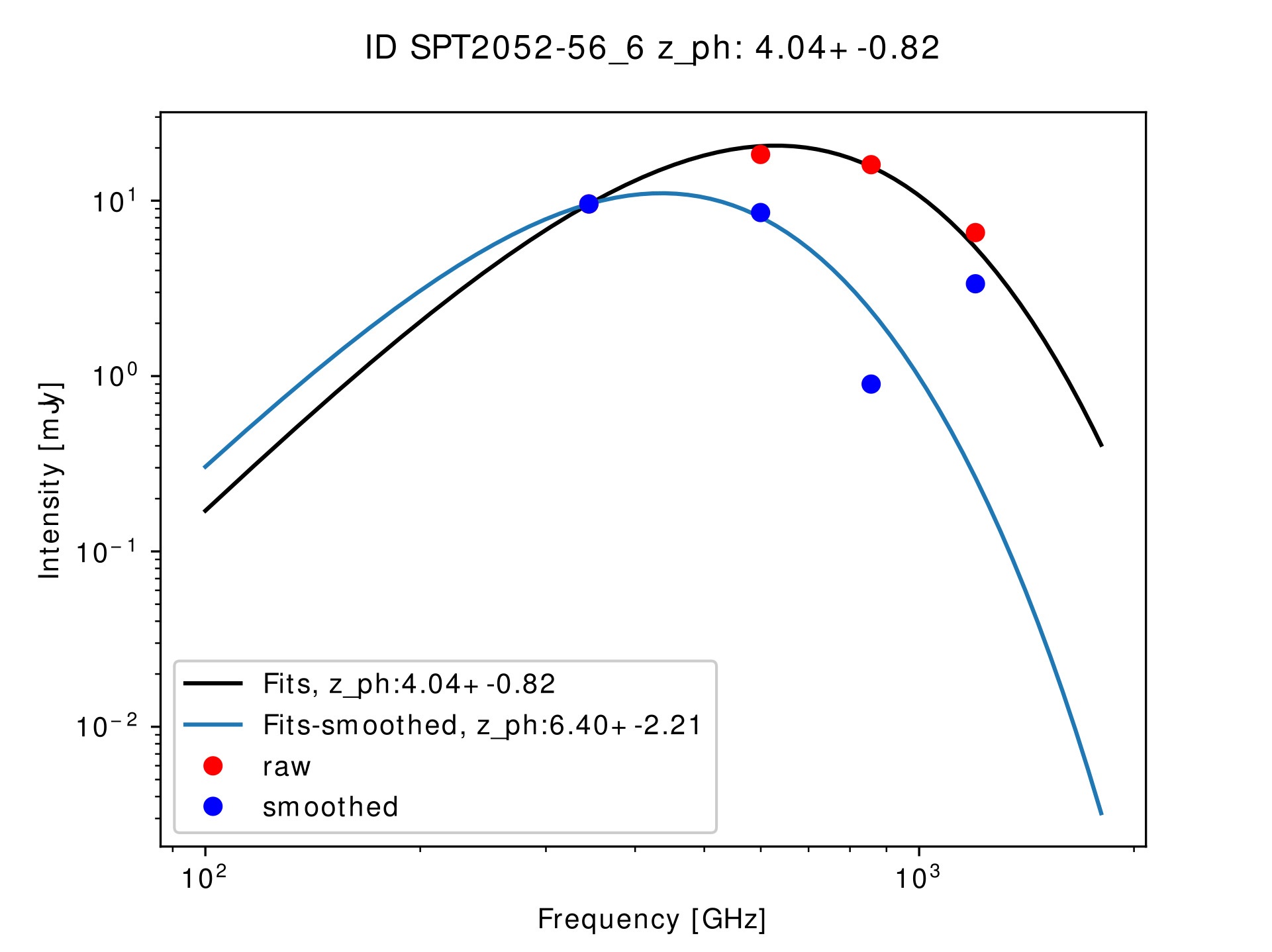}}
	\subfloat[\texttt{ID SPT0311-58\_1}, $z_{ph}: 5.29\pm1.36$]{\label{fg3}\includegraphics[width=8cm]{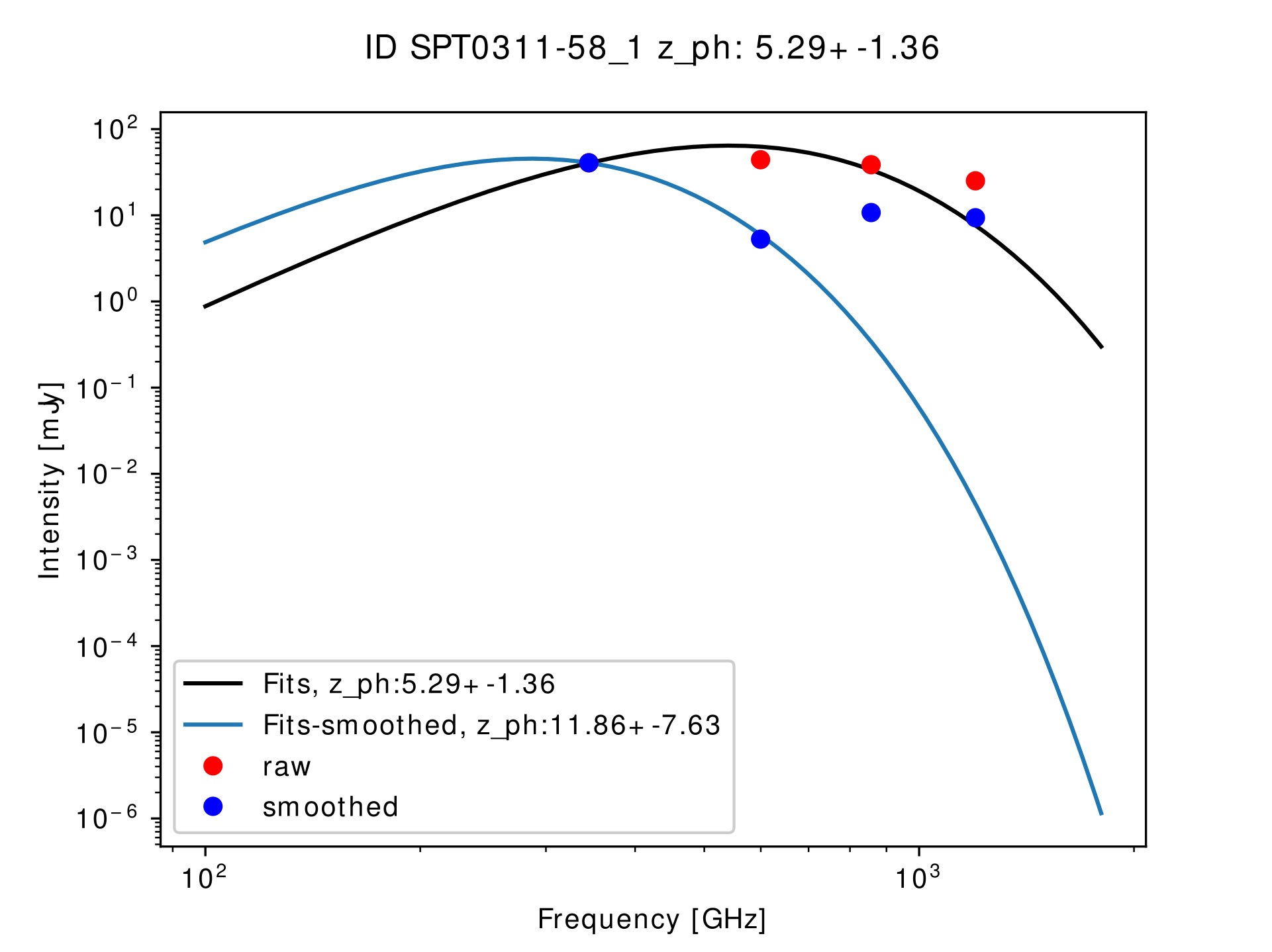}}
	\label{figur}\caption{Example graphs which original fits gives better result than smoothed version}
\end{figure*}

\section{3. Conclusion and Outlook}
This research focuses on the extracting source data form the LABOCA(LArge APEX BOlometer CAmera) and SPIRE(The Spectral and Photometric Imaging Receiver) data, and specify the coordinate of the source and calculate their red-shift values. 

We developed the algorithm using Multi-Gaussian fitting, which primary coding environment was Python, and for the data smoothing, we used MIRIAD. 
We found the source information, including their coordinates(RA, DEC), Intensity(mJy), the uncertainty of each parameter, red-shift from the original LABOCA and SPIRE(SPIRE image) map by using multiple Gaussian fitting Method. 
Also, a large part of the LABOCA maps are empty and do not show astronomical signals, as we can see in Fig 1.
Thus first, we use the "boxing method" for bounding the analysis range, which has dense source density compare to the void region. This improves the Gaussian fitting accuracy because only the relevant parts of a map are processed 
by the code.

And then, one needs to decompose each box into individual galaxies. Moreover, for the fitting algorithm for one-source is in \textbf{Algorithm. 2} 
Moreover, after that, use the Gaussian fitting for each box. 
To calculate the number of peaks in one source, we used to sum up the number of Gaussian peaks. The fitting the sum of single Gaussian to the source coordinate is only the first step to derive the number of Gaussian components and a guess for their location and intensity. The FWHM of the Gaussian is fixed with a small margin to the beamwidth of the map that this decomposition of the source structure is then used as an initial guess for the simultaneous fitting of all gaussian components.

Cross-matching the galaxies identified in the LABOCA map in the SPIRE bands is a multi-step process which is first comparing the coordinates in an individual tolerance, second fitting a gaussian at the Laboca position on the SPIRE map and last integrating the pixel fluxes at the LABOCA position in case the
Gaussian fit does not give reliable results. 
Because the source may not show up in the SPIRE maps (in particular at 350 or 250um) and verify that the flux densities are extracted belong to the same object. 

Moreover, after running the reshift calculation, we can compare the original las smoothed result.
For the smoothed result, Smoothed data is fully better at \texttt{SPT0311-58 : \_9, \_5}, \texttt{SPT2052-56 : \_7, \_8}. And smoothed data is partially better source was, \texttt{SPT0311-58 : \_3, \_7}, \texttt{SPT0348-62 : \_5}, \texttt{SPT2018-45 : \_5, \_6, \_7, \_8}

In these studies, we did not evaluate all data points critically. However, one could try to improve the extraction of the source fluxes by using the resulting dust SED fits in a more systematic way to identify outliers to improve the flux fitting in these cases.

As a result, we analyze the 250, 350, 500, 850 $\mu m$ wavelengths image map and calculate each source red-shift value with $\pm0.5$ uncertainty for good cases.

\section{Acknowledgement}
I would like to thank Dr. Axel Weiß for inspiring discussions and also appreciate Prof. Karl M. Menten for the opportunity to work on this project. Also thanks for the members of MPIfR for the valuable comments to the draft.

\section{Appendix}

\subsection{A.1 Other LABOCA image examples}

Other examples for LABOCA data \cite{3} \cite{4} 
In Fig. 11, once can see (\texttt{Laboca/SPT2052-56-model.fits}). Fig. 11 (b) One can see the decomposed bright peak from the flux map. In Fig. 11 (c) one can see five bright peaks. Fig. 11 (d) Compare to Fig. 11 (c) one can see the whole map become more flattered without peaks.

\begin{figure*}[htp]
	\centering
	\subfloat[The global flux map, LABOCA. ]{\label{figur:1}\includegraphics[width=85mm]{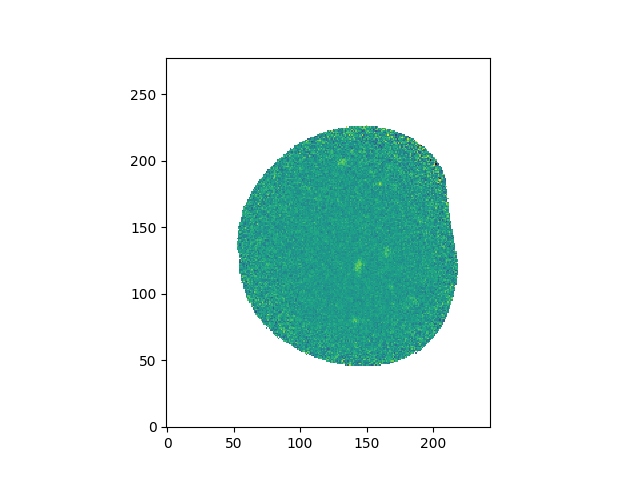}}    
	\subfloat[The flux map, LABOCA. ]{\label{figur:2}\includegraphics[width=85mm]{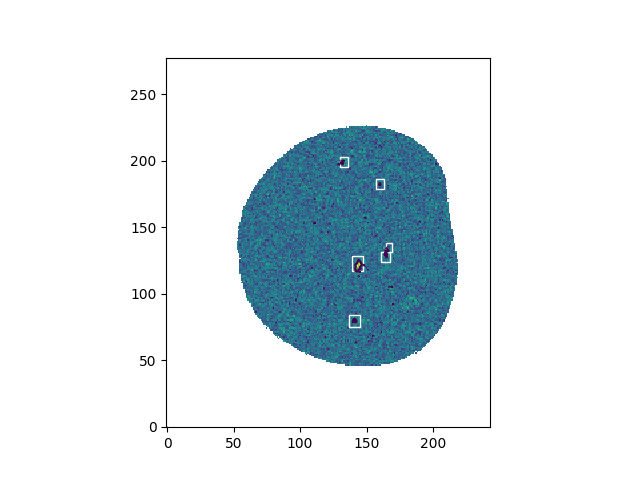}}
	\\
	\subfloat[The fitted map, LABOCA. ]{\label{figur:3}\includegraphics[width=85mm]{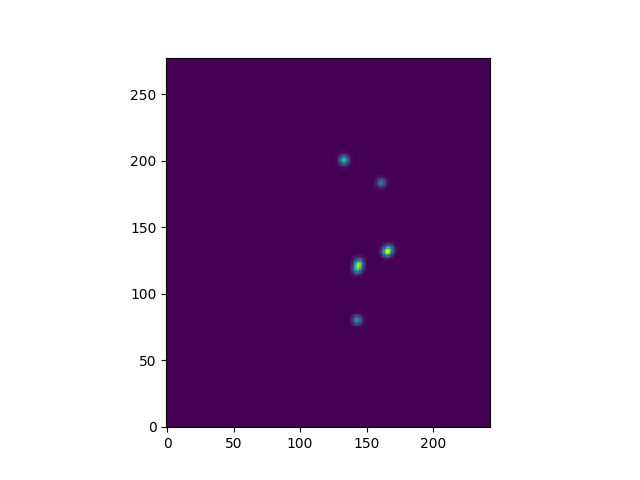}}    
	\subfloat[The residual map. ]{\label{figur:4}\includegraphics[width=85mm]{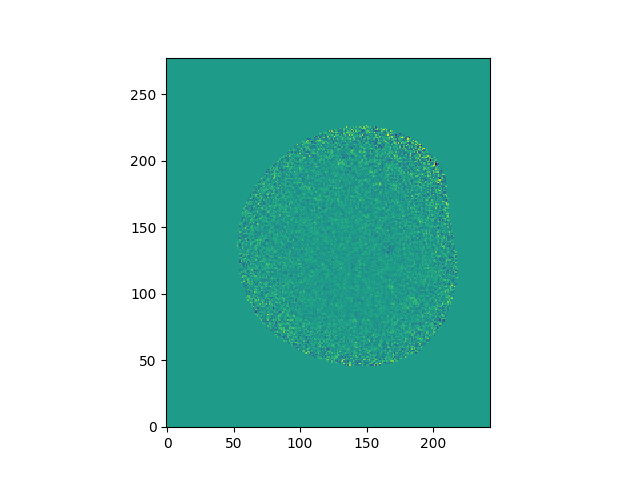}}
	
	\label{figur}\caption{(a), (b) One needs to which we aim to decompose into the bright peak from the flux map. (c) One can see five bright peaks. (d) Compare to Fig. 11 (c) one can see the whole map become more flattered without peaks.}    
\end{figure*}

\subsection{A.2 SPIRE data smoothing process with \texttt{MIRIAD}}

\begin{lstlisting}[language=Python]
>> execfile('source-extractor-version2.0.py')
>> dividefits()
>> getmx()
\end{lstlisting}

The \texttt{MIRIAD} is the radio-astronomy data reduction package used by the ATNF and has been ported to a variety of UNIX-based platforms. 
The first process is dividing the \texttt{.fits} files by its layer data. And record its data as \texttt{.mx} file which can use it at \texttt{MIRIAD}.

\begin{breakablealgorithm}
	\caption{SPIRE data layer divide code}
\begin{lstlisting}[linewidth=\columnwidth,breaklines=true,language=Python]
def dividefits():
	inlist=['SPIRE/SPT2335-53-SPIRE250.fits',
	'SPIRE/SPT2335-53-SPIRE500.fits',
	'SPIRE/SPT2335-53-SPIRE350.fits',
	'SPIRE/SPT0303-59-SPIRE250.fits',
	'SPIRE/SPT0303-59-SPIRE350.fits',
	'SPIRE/SPT0348-62-SPIRE250.fits',
	'SPIRE/SPT0348-62-SPIRE350.fits',
	...
	'SPIRE/SPT2018-45-SPIRE500.fits',
	'SPIRE/SPT2052-56-SPIRE250.fits',
	'SPIRE/SPT2052-56-SPIRE350.fits',
	'SPIRE/SPT2052-56-SPIRE500.fits']    

	for file in inlist:
	print file
	convertfits(infile=file)
\end{lstlisting}
\end{breakablealgorithm}

\begin{breakablealgorithm}
	\caption{\texttt{MIRIAD} script producing code}
	\begin{lstlisting}[linewidth=\columnwidth,breaklines=true,language=Python]
def getmx():
	band=['250', '350', '500']
	f1=open("SPIRE/miriad.mx",'w')
	for j in range(3):
	for i in range(8):
	infile  = "SPIRE/"+sourcelist[i]+"-SPIRE"+band[j]+"-flux.fits"
	print infile 
	filename = "smooth-"+sourcelist[i]+"-SPIRE"+band[j]+"-flux.mx" 
	f1.write("source %s\n"%filename)
	hdulist=fits.open(infile)
	header=hdulist[0].header

	########  smooth.mx  #########
	f=open("SPIRE/"+filename,'w')
	f.write("rm -rf flux\nrm -rf rms\nrm -rf flux-reg\nrm -rf rms-reg\nrm -rf flux-sm\nrm -rf rms-sm\nrm -rf rms1\n")

	filename = sourcelist[i]+"-SPIRE"+band[j]+"-flux.fits" 
	f.write("fits in=%s out=flux op=xyin\n"%(filename))

	filename = sourcelist[i]+"-SPIRE"+band[j]+"-rms.fits" 
	f.write("fits in=%s out=rms op=xyin\n"%filename)
	f.write("itemize in=flux\nprthd in=flux\n")
	f.write("regrid in=flux out=flux-reg axes=1,2 desc=%2.10f,%i,%2.10fe-05,%i,%2.10f,%i,%2.10fe-05,%i options=nearest\n"%(header['CRVAL1']*math.pi/180, 2*header['CRPIX1'], 100000*header['CDELT1']*math.pi/360, 2*header['NAXIS1'], header['CRVAL2']*math.pi/180, 2*header['CRPIX2'], 100000*header['CDELT2']*math.pi/360, 2*header['NAXIS2']))
	f.write("regrid in=rms  out=rms-reg  axes=1,2 tin=flux-reg options=nearest\n")
	if '250' in infile:
	fwhm=7.16
	if '350' in infile:
	fwhm=9.72
	if '500' in infile:
	fwhm=14.31
	f.write("smooth in=flux-reg out=flux-sm fwhm=%2.2f\n"%fwhm)
	f.write("smooth in=rms-reg  out=rms1    fwhm=%2.2f scale=0.0\n"%fwhm)
	R=(1/1.0795)**2
	f.write("prthd in=flux-sm\nmaths 'exp=<rms1>*%2.4f' out=rms-sm\n"%R)

	filename = sourcelist[i]+"-SPIRE"+band[j]+"-flux-sm.fits" 
	f.write("fits in=flux-sm out=%s op=xyout\n"%filename)

	filename = sourcelist[i]+"-SPIRE"+band[j]+"-rms-sm.fits" 
	f.write("fits in=rms-sm out=%s op=xyout\n"%filename)
	f.write("rm -rf flux\nrm -rf rms\nrm -rf flux-reg\nrm -rf rms-reg\nrm -rf flux-sm\nrm -rf rms-sm\nrm -rf rms1")
	f.close()
	f1.close()
\end{lstlisting}
\end{breakablealgorithm}

\begin{lstlisting}[language=Python]
% source miriad.mx
\end{lstlisting}

After that we can run \texttt{miriad.mx} file for smoothing.

\begin{lstlisting}[language=Python]
>> mergefits()
\end{lstlisting}

After finishing the smoothing, we can merge the \texttt{.fits} files into the original form before \texttt{divideout}.

\begin{breakablealgorithm}
	\caption{SPIRE map layers after smoothing; \texttt{mergefits()} code}
\begin{lstlisting}[linewidth=\columnwidth,breaklines=true,language=Python]
def mergefits():
band=['250', '350', '500']
for j in range(3):
for i in range(8):
infile  = "SPIRE/"+sourcelist[i]+"-SPIRE"+band[j]+".fits"
print infile 
infile1 = "SPIRE/"+sourcelist[i]+"-SPIRE"+band[j]+"-flux-sm.fits"
infile2 = "SPIRE/"+sourcelist[i]+"-SPIRE"+band[j]+"-rms-sm.fits"
infile3 = "SPIRE/"+sourcelist[i]+"-SPIRE"+band[j]+"-snr.fits"
hdulist=fits.open(infile)
hdulist1=fits.open(infile1)
hdulist2=fits.open(infile2)

hdulist[1].data=copy.deepcopy(hdulist1[0].data)
hdulist[2].data=copy.deepcopy(hdulist2[0].data)

if len(str(infile).split('/')) > 1:
outfileflux="SPIRE/"+str(infile).split('/')[1].split('.')[0]+"-sm.fits"
else:
outfileflux="SPIRE/"+str(infile).split('.')[0]+"-sm.fits"

com='rm -rf '+outfileflux
os.system(com)
hdulist.writeto(outfileflux)
\end{lstlisting}
\end{breakablealgorithm}

\end{document}